\documentstyle[aps]{revtex}
\begin{document}

\title{Finite cutoff on the string worldsheet?}
\author{Vipul Periwal and \O yvind Tafjord}
\address{Department of
Physics,
Princeton University,
Princeton, New Jersey 08544}

\def\dd{\hbox{d}}
\def\tr{\hbox{tr}}\def\Tr{\hbox{Tr}}
\def\ee#1{{\rm e}^{{#1}}}
\def\Bp{\left|B_p\right>}
\maketitle
\begin{abstract}
D-brane backgrounds are specified in closed string theories by
holes with appropriate mixed Dirichlet and Neumann  boundary conditions
on the string worldsheet.  As presently stated, the prescription
defining D-brane backgrounds is such that the Einstein equation is not
equivalent to the condition for scale invariance on the string
worldsheet.  A modified D-brane prescription is found, that leads to
the desired equivalence, while preserving all known D-brane lore.
A possible interpretation is that the worldsheet cutoff is finite.
Possible connections to recent  work of Maldacena and Strominger,
and Gopakumar and Vafa are suggested.
\end{abstract}

\medskip

\section{Introduction}

One of the basic successes of the framework of string theory is the
consistent way in which one can introduce background condensates of
fields. Only certain configurations of background fields are allowed
by conformal invariance on the string worldsheet. It is crucial for
the consistency of string propagation that conformal invariance on
the worldsheet is maintained---it implies unitarity, for example. A
primary reason
for thinking that string theories have something to do with
fundamental physics, is that the conditions for scale invariance on
the world-sheet turn out to be equivalent to physically interesting
equations of motion for
the spacetime metric and other fields\cite{ft,cfmp}.

For instance, consider bosonic closed string theory with a background
metric $G_{\mu\nu}(X)$ and dilaton field $\Phi(X)$ (for simplicity
defined with zero expectation value). This is described by a sigma
model action (we use notation and conventions as in
Polchinski\cite{polchinski})
\begin{equation}
S={1\over4\pi\alpha'}\int d^2\sigma\sqrt{g}\left[
g^{ab} G_{\mu\nu}(X)\partial_a
X^\mu\partial_b X^\nu+\alpha' R\Phi(X)\right].
\end{equation}

Conformal invariance at
the quantum level corresponds to requiring the beta functions to
vanish. At one loop this means
\begin{eqnarray}\label{betas}
\beta_{\mu\nu}&=&\alpha'R_{\mu\nu}+2\alpha'\nabla_\mu\nabla_\nu\Phi=0,\\
\beta_{\Phi}&=&-{\alpha'\over2}\nabla^2\Phi+\alpha'(\nabla\Phi)^2=0. \nonumber
\end{eqnarray}
These equations are equivalent to the equations of motion following
from the spacetime action
\begin{equation}
S^{\rm closed}_{\rm eff}={1\over 2\kappa^2}\int
d^Dx\sqrt{-G}e^{-2\Phi}\left[R+4(\nabla\Phi)^2\right],
\end{equation}
where we are dropping terms that are zero in the relevant case of $D=26$.
Similarly, if we add open strings and a gauge field background on the
boundary, we obtain conditions for conformal invariance equivalent to
the equations of motion for a spacetime Born-Infeld action including
metric and dilaton contributions, $S_{\rm eff}^{\rm open}$. Thus the
total spacetime action is given by $S^{\rm closed}_{\rm eff}+cS^{\rm
open}_{\rm eff}$. The constant $c$ is determined by the coupling
between gravity and gauge fields, and it appears, for instance,
multiplying the gauge field energy-momentum tensor in the full
Einstein equation.

The closed string beta functions (\ref{betas}),
which give the
vacuum Einstein equation, do not change if one adds a boundary to
the worldsheet, but the full equation is nevertheless obtained by a
generalized form of conformal invariance including the effects of
worldsheets of
different topologies. This all works due to the Fischler-Susskind
mechanism\cite{FS} where the extra breaking of conformal invariance
comes from shrinking fixtures on the worldsheets of higher genus. In the
present case it arises when the disk degenerates to a sphere, i.e.,
we represent the disk as a sphere with a hole cut out, and upon
integrating the size of the hole there will be a logarithmic
divergence for small hole size. The coefficient of the divergence
looks like an insertion of a local operator on the sphere.
Identifying the cutoff on the hole size with the cutoff used for the
counterterms in the sphere amplitude, we can demand cutoff
independence of the combined amplitude. This yields the full Einstein
equation, and thus determines the constant $c$ \cite{FS,callan}.

The case of a D-brane background\cite{green} is in many ways
analogous to the
gauge field background, in fact, the latter is a special case (a
D25-brane in the bosonic case). A static flat D$p$-brane is
introduced simply by adding a boundary on the worldsheet with Neumann
boundary conditions on $p+1$ of the directions and Dirichlet boundary
conditions on the others. The sphere amplitude gives the same beta
functions as above (\ref{betas}), while the disk amplitude gives
conditions corresponding to equations of motion following from a
Dirac-Born-Infeld action\cite{leigh} (which in general also have
contributions from the background gauge field and antisymmetric tensor field)
\begin{equation}
S_{\rm eff}^{\rm DBI}=c\int d^{p+1}\xi e^{-\Phi}\sqrt{-\tilde{G}},
\end{equation}
where $\xi^A,\ A=0,\ldots,p$ are coordinates on the D-brane world
volume, and $\tilde{G}_{AB}$ is the induced metric.

The full spacetime effective action should therefore be given by
$S_{\rm eff}^{\rm closed}+S_{\rm eff}^{\rm DBI},$ and the
corresponding Einstein equation must have a source term due to the
D-brane. In analogy with the open string case above, it would be
natural to expect this source term to appear from the worldsheet
point of view as a Fischler-Susskind\cite{FS} type effect. Thus
one expects that there is a logarithmic divergence in the disk
amplitude, whose coefficient represents the effect of the D-brane
in the Einstein equation\cite{schmid}. We would then have to
correct the sphere metric to one that solves these equations to
compute consistent amplitudes. This contradicts the prescription
for introducing D-branes into the theory, which says that one can
either expand about background fields representing the
(super-)gravity solution of the soliton (and thus extract
restricted long-distance physics), or one can use the remarkable
prescription of adding boundaries with Dirichlet boundary
conditions\cite{green}, using a flat background metric. We should
not need to do both! Indeed, explicit calculations involving
D-branes show no logarithmic divergence in the disk amplitude for
localized ($p+1<D$) branes\cite{larus,rob}. Furthermore,
Leigh\cite{leigh} explicitly noted that the disk amplitude does
not change the closed string beta functions. The aim of this paper
is to resolve this puzzle.

The two ways of introducing D-branes mentioned above are analogous to
two different interpretations of the spacetime effective action, where
one can either expand supergravity about a vacuum corresponding to the
D-brane supergravity solution, or couple supergravity to a
world-volume action (the DBI action) representing the D-brane.  These
two descriptions are useful in somewhat complementary regimes, the
latter for isolated D-branes, and the former for $N$ superposed
D-branes, with $Ng_{{\rm st}}$ not necessarily small.  We also note
that a fact that does separate the case of localized branes from the
open string case is that the Einstein equation imply that the beta
functions are set to zero everywhere except at the source.

\section{The bosonic case}

Let us look at the disk amplitude in more detail. For concreteness we
will consider the two graviton amplitude in the background of a
D$p$-brane at the origin, first computed in \cite{larus}, but the
discussion is general.
To isolate the part of the amplitude associated with the small hole,
we represent the disk as a sphere with a hole cut out and integrate
also over the radius $a$ and position $z$ of the hole. This increases
the symmetry of the amplitude from SL(2,${\bf R}$) to SL(2,${\bf C}$)
as in the sphere amplitude. The amplitude can be written\cite{rey,FKS}
\begin{equation}
A_{\rm disk}=2\pi^2\tau_p\int2{da\over
a^3}\int_{|z_\alpha-z|>a}\!\!\!d^2z\prod_{\alpha}d^2 z_\alpha{1\over
V_{\rm CKG}}\left< \prod_\alpha V_{\alpha}(z_\alpha)\right>_{D_2}.
\end{equation}
Here $2\pi^2\tau_p$ is the correct normalization for the D$p$-brane
disk amplitude, with $\tau_p$ being the D$p$-brane
tension\cite{polchinski}. We have
$V_\alpha(z_\alpha)={\kappa\over\pi\alpha'}
\epsilon_{\mu\nu}^\alpha\partial X^\mu\bar{\partial}X^\nu
e^{ik_\alpha\cdot X}(z_\alpha)$ for gravitons, and $V_{\rm CKG}$ denote the
volume of the conformal killing group which we need to divide out.
The $X^\mu(z)$ satisfy boundary conditions appropriate for a
D$p$-brane \begin{eqnarray}
\partial_n X^I(z)|_{\partial\Sigma}&=&0,\ \ I=0,\ldots, p,\\
X^i(z)|_{\partial\Sigma}&=&0,\ \ i=p+1,\ldots,D-1.
\end{eqnarray}
We could easily have included the effects of a constant field
strength background $F_{AB}$. This would simply modify the
boundary conditions and Green functions as in~\cite{callan}, and
the rest of the discussion would go through as below, with
identical conclusions. The
exact Green functions for this geometry are for Neumann and Dirichlet
directions respectively,
\begin{eqnarray}
\left<X^{J}(z_1)X^I(z_2)\right>&=&{\alpha'\over2} \eta^{IJ}\left(
-\ln|z_1-z_2|^2-\ln\left|1-{a^2\over(z_1-z)(\bar{z}_2-\bar{z})}\right|^2\right),
\\
\left<X^{j}(z_1)X^i(z_2)\right>&=&{\alpha'\over2} \eta^{ij}\left(
-\ln|z_1-z_2|^2+\ln\left|1-{a^2\over(z_1-z)(\bar{z}_2-\bar{z})}\right|^2
\right.+\nonumber\\
& &\ \ +\left.\ln\left|{(z_1-z)(z_2-z)\over a}\right|^2\right). \label{GD}
\end{eqnarray}
Note that the last term in (\ref{GD}) must be kept even if it factorizes,
due to the lack of momentum conservation in the transverse
directions. We can now perform the contractions in $A_{\rm disk}$ and
expand the integrand for small $a$. We find
\begin{eqnarray}\label{disk}
A_{\rm disk}&=&2\pi^2\tau_p\int2{da\over
a^3}\int_{|z_\alpha-z|>a}\!\!\!\!\!\!\!d^2z\prod_{\alpha}d^2 z_\alpha
{1\over V_{\rm CKG}}
a^{{1\over2}\alpha'k_{\perp}^2}\times\nonumber\\
& &\times\left[\left<T(-k_\perp,z)\prod_\alpha
V_{\alpha}(z_\alpha) \right>_{S_2}
{+}a^2\left<D(-k_\perp,z)\prod_\alpha
V_{\alpha}(z_\alpha)\right>_{S_2}{+}{\cal O}(a^4)\right],
\end{eqnarray}
where $k^i_\perp$ denotes the total transverse momentum $\sum_\alpha
k_\alpha^i,$ and $T(k,z)={:}e^{ik\cdot X}(z){:}$ and
$D(k,z)={2\over\alpha'} (\eta M)_{\mu\nu}{:}\partial
X^\mu\bar{\partial}X^\nu e^{ik\cdot X}(z){:}$ are (off-shell) tachyon
and graviton/dilaton vertex operators (not properly normalized for
simplicity). The flat metric is $\eta_{\mu\nu}={\rm
diag}(-1,1,1,\ldots,1)$ while $M_{\mu\nu}\equiv{\rm
diag}(1,\ldots,1,-1,\ldots,-1)$ with signature $(p+1,D-p-1)$. We
denote expectation values on the sphere and disk by $\left<\
\right>_{S_2}$ and $\left<\ \right>_{D_2}$ respectively. In the sphere
amplitudes we have only kept the zero mode integration in the Neumann
direction, giving rise to a factor $(2\pi)^{p+1}\delta^{p+1}(\sum
k_\parallel)$.

For the special case of open strings (space-filling brane) we have
$k_\perp=0, $ and the second term above gives rise to a logarithmic
divergence in the hole size cutoff, proportional to a zero-momentum
dilaton. This is the Fischler-Susskind mechanism yielding a
cosmological constant in the Einstein equation\cite{FS}. However, for
localized D-branes there is an extra factor of
$a^{\alpha'k_\perp^2/2},$ and integrating over $a$ gives rise to poles
in $k_\perp^2, $ rather than divergences. In the case of graviton
scattering off the D-brane, these are physical $t$-channel poles. Thus
we reproduce the well-known result that the disk amplitude is finite,
and there is no need to introduce a cutoff on the hole size. A
spacetime effective action can be deduced by examining the poles in
these well-defined disk amplitudes\cite{rob}. Thus, if we were not
interested in seeing how the full Einstein equation is related to
conformal invariance on the worldsheet, we would stop here and be
content with this answer for the scattering amplitude.

Let us now examine the full equations of motion following from the
spacetime effective action $S_{\rm eff}^{\rm closed}+S_{\rm
eff}^{\rm DBI}.$ Consider a flat, static D$p$-brane, then the metric
and dilaton equations of motion can be written
\begin{eqnarray}
{e^{-2\Phi}\over2\kappa^2\alpha'}\beta_{\mu\nu}&=&
{c\over4} e^{-\Phi}(\eta
M)_{\mu\nu}\delta^l(x_{\perp})\label{einst1}\\
{e^{-2\Phi}\over2\kappa^2\alpha'}(8\beta_\Phi-2\beta^\mu_\mu)&=&
ce^{-\Phi}\delta^l(x_{\perp}).
\end{eqnarray}
In these equations we have already expanded the left hand sides for a
flat background metric. We
have defined $l=D-p-1$, the number of transverse directions. To
illustrate our point, let us for the moment concentrate on the case
$p=11$ (and $D=26$) where it is consistent to choose $\Phi=0.$
It is then
easy to linearize by setting $G_{\mu\nu}=\eta_{\mu\nu}+h_{\mu\nu}$
and solve for the leading term in $h_{\mu\nu}$. In Hilbert gauge we
have $R_{\mu\nu}=-{1\over2}\partial^2 h_{\mu\nu},$ and we get (for
$l=14$)
\begin{equation}\label{metric}
h_{\mu\nu}=\kappa^2 c(\eta
M)_{\mu\nu}\int{d^lq_{\perp}\over (2\pi)^l}{1\over
q_\perp^2}e^{iq_\perp\cdot x_{\perp}}.
\end{equation}
We now
investigate what we get when we insert this metric perturbation
on the sphere. To leading order the effect is an insertion of
$h_{\mu\nu}\partial X^\mu\bar{\partial}X^\nu$ into the amplitude. We
normal order this operator using
$e^{iqx}=\epsilon^{\alpha'q^2/2}{:}e^{iqx}{:}$
where $\epsilon$ is the cutoff on the worldsheet. In addition, there
are contractions between the exponential in $h_{\mu\nu}$ and
$\partial X^\mu\bar{\partial}X^\nu.$ The term with both derivatives
contracted contributes to the tachyon pole, and will be neglected
together with similar terms below. The terms with one derivative
contracted give rise to a total derivative, while contracting the
derivatives with each other gives worldsheet curvature terms
important for the dilaton equation of motion, as discussed in
\cite{FKS,alwpol}. We will also not worry about these for our
purposes as we concentrate on the $\beta_{\mu\nu}$ equation of
motion.

Taking this into account, the leading order effect of using
the metric (\ref{metric}) in the sphere amplitude is given by
\begin{equation}\label{sphere}
A_{\rm sphere}^{(1)}=-{32\pi^3\over\alpha'\kappa^2}\cdot
{1\over2\pi\alpha'}\cdot\kappa^2 c\cdot{\alpha'\over2}\cdot
{\epsilon^{{1\over2}\alpha'k_\perp^2}\over k_\perp^2}\int
d^2z\prod_\alpha d^2z_\alpha \left<D(-k_\perp,z)\prod_\alpha
V_\alpha(z_\alpha) \right>_{S_2}.
\end{equation}
Here the two first terms come from the normalization of the sphere
amplitude and the prefactor of the action, respectively. The second
and third terms come from $h_{\mu\nu}$ and the normalization of
$D(-k_\perp,z)$. The momentum conservation in the Dirichlet directions
has been used to integrate over $q_\perp$ (so again only the zero
modes in the Neumann directions remain).  In the background of a
non-trivial metric, we should also correct the vertex operators so
that they have the correct conformal weights with respect to the new
energy-momentum tensor. However, for the two-graviton amplitude we
consider here, the effect to this order will look like a two-point
sphere amplitude and thus vanish, so we can use the vertex operators
for the flat metric.

Now we compare this to the disk amplitude (\ref{disk}). We neglect
the tachyon pole and focus on the massless pole. One obvious
contribution comes from using the second term in the $a$ expansion
while neglecting the higher order effects of the integration domain.
We can then integrate over $a$, with a lower limit $\epsilon$, and we
find that the $\epsilon$-dependent part exactly cancels the
$\epsilon$ dependence of the sphere amplitude (\ref{sphere}) provided
we choose \begin{equation}\label{ceq}
c=-\tau_p,
\end{equation}
precisely as expected! We only did this for
$p=11$, but, as we will see below, it holds for all $p$.
Alternatively, we can say that the contribution from the sphere, with
cutoff $\epsilon$, exactly compensates for the piece left out from
the disk amplitude by cutting off the hole size at $\epsilon$,
provided we have ``synchronized'' the sphere and the disk by choosing
the correct metric, i.e., solved the correct Einstein equation with
source term and specific value of $c$. We can take $\epsilon$ to zero
if we want, then the sphere contribution vanishes, and the disk gives
the full answer. If we want to extract a sensible long distance
answer from the sphere by itself, as we would expect to be able to
do, we will have to fix a non-zero $\epsilon$ and look at processes
of very low momentum transfer,
\begin{equation}
    k_\perp^2\ll{2\over\alpha}{1\over|\ln\epsilon|}.
    \label{length}
\end{equation}
Then
$\epsilon^{\alpha'k_\perp^2/2}=1+{1\over2}\alpha'k_\perp^2\ln\epsilon+\ldots,
$ 
$\epsilon$-independent amplitude which corresponds to the
semiclassical approximation of the amplitude, coinciding with the
amplitude found from Polchinski's prescription in the same limit.

There are also other contributions to the massless pole in the disk
amplitude. One comes from the tachyon insertion, considering ${\cal
O}(a^2)$ effects of the integration region. Another comes from
certain parts of the ${\cal O}(a^4)$ term in the expansion, for which
integration over $z$ yields extra $1/a^2$ singularities. These two
contributions should cancel as discussed in \cite{FKS}.

A more formal way of deriving the same results is to consider the
sphere amplitude with an arbitrary metric. We then need to introduce
a cutoff, and at one loop we must put in a counterterm of the type
$\beta_{\mu\nu}\partial X^\mu\bar{\partial}X^\nu\ln\epsilon.$ This is
formal because the metrics involved here are not so smooth as to
allow the normal coordinate expansion employed in deriving this.
Nevertheless we can consider the sum of the sphere and the disk
amplitude and ask for cutoff independence in some sense. Neglecting
the tachyon pole, we want to demand that
$\epsilon\partial_\epsilon(A_{\rm sphere}+A_{\rm disk})$ vanishes to
leading order in $\epsilon$ (note that away from $k_\perp=0$ this
derivative always vanishes for the disk amplitude as we take
$\epsilon$ to zero). This will precisely be true if $\beta_{\mu\nu}$
is given as in (\ref{einst1}), with $c$ given in (\ref{ceq}), using
the same normal ordering considerations as we did above (treating
$\beta_{\mu\nu}$ as $\epsilon$-independent when taking the
derivative). This holds for any $p$. Thus we have a way of extracting
the full Einstein equation---not by demanding conformal invariance,
which we already have on the disk alone---but by introducing a cutoff
anyway and demanding agreement between the contributions from the
sphere and the disk, cancelling the {\it leading order} cutoff
dependence (neglecting tachyon effects) even if the dependence
vanishes as the cutoff goes to zero.

\section{The superstring case}

Let us now do the same calculation in the case of the type II
superstring. In this case we want to reproduce the full supergravity
equations of motion, including a source term for the D$p$-brane. The
source term should again arise from the disk-amplitude, while the rest
of the equation is given in terms of the $\beta$-functions on the
sphere and involve the metric, the dilaton, and the appropriate
Ramond-Ramond gauge field. These equations are well known, as are the
D$p$-brane solutions. For clarity, we again concentrate on the
constant dilaton example, the D3-brane. For our purposes we simply
need the long distance behavior of the metric, conveniently given by
Garousi and Meyers\cite{rob} as
\begin{equation}
h_{\mu\nu}={\kappa^2\tau_3\over4\pi^3 x_\perp^4}(\eta M)_{\mu\nu}=
\kappa^2\tau_3(\eta
M)_{\mu\nu}\int{d^6q_{\perp}\over (2\pi)^6}{1\over
q_\perp^2}e^{iq_\perp\cdot x_{\perp}}.
\end{equation}
When this solution is inserted in the sphere amplitude, there will be
a term of the following form,
\begin{equation}\label{Asph}
A_{\rm sphere}\sim-{32\pi^3\over\alpha'\kappa^2}\cdot
{1\over2\pi\alpha'}\cdot\kappa^2\tau_3\cdot{\alpha'\over2}\cdot
{\epsilon^{{1\over2}\alpha'k_\perp^2}\over k_\perp^2}\int
d^2z\prod_\alpha d^2z_\alpha \left<D(-k_\perp,z)\prod_\alpha
V_\alpha(z_\alpha) \right>_{S_2}.
\end{equation}
There will be other terms, from the world sheet fermions and the
Ramond-Ramond gauge field, but these will follow from this term by
world sheet and space time supersymmetry, since the D$p$-brane is a
BPS state.  This amplitude has all the same features as the
corresponding bosonic amplitude, and we now want to show that the disk
amplitude reproduces this behavior, again in full analogy with the
bosonic case. To evaluate the superstring disk amplitude, we find it
convenient to use the boundary state formalism\cite{callanbound,li},
where the D$p$-brane is represented by a closed string boundary state,
\begin{equation}
\Bp=4\cdot2\pi^2\tau_p\cdot
\delta^l(x_\perp)\exp\left[-\sum_{n=1}^\infty{1\over n}
(\eta M)_{\mu\nu}\tilde{\alpha}_{-n}^\mu\alpha_{-n}^\nu-
\sum_{r={1\over2},{3\over2},\ldots}(\eta M)_{\mu\nu}
\tilde{\psi}_{-r}^\mu\psi_{-r}^\nu\right]\left|0\right>.
\end{equation}
There is also a ghost piece in $\Bp$ which will not be important
here. The normalization is chosen such that amplitudes calculated
using $\Bp$ agree with the corresponding disk amplitudes. In closed
string operator language, the disk amplitude then reads
\begin{equation}
A_{\rm disk}=\left<V\right|V\Delta V\cdots \Delta\Bp,
\end{equation}
where the propagator is
\begin{equation}
\Delta={1\over2}(L_0+\tilde{L}_0-1)^{-1}\longrightarrow
{1\over4\pi}\int{d^2z\over|z|^2}z^{L_0-{1\over2}}
\bar{z}^{\tilde{L}_0-{1\over2}},
\end{equation}
where the last expression also includes a projection over states
annihilated by $L_0-\tilde{L}_0$ (as is the case for physical
states). For completeness, we give
\begin{equation}
L_0={\alpha'\over4}p^2+\sum_{n=1}^\infty\alpha_{-n}^\mu\alpha_{n\mu}
+\sum_{r={1\over2},{3\over2},\ldots}
r\psi_{-r}^\mu\psi_{r\mu},
\end{equation}
with a similar expression for $\tilde{L}_0$. We can then derive the key result
\begin{eqnarray}
\Delta\Bp&=&4\cdot2\pi^2\tau_p\cdot{1\over4\pi}\int{d^2z\over|z|^2}
|z|^{{1\over2}\alpha' k_\perp^2}\prod_{n,r}\times\nonumber\\
& &\ \ \exp\left[-|z|^{2n}(\eta M)_{\mu\nu}{1\over n}\tilde{\alpha}_{-n}^\mu
\alpha_{-n}^\nu-|z|^{2r}(\eta M)_{\mu\nu}\tilde{\psi}_{-r}^\mu
\psi_{-r}^\nu\right]\left|\,0;-\sum k_\perp\right>\nonumber\\
&=&4\pi^2\tau_p\int{da\over a^2}a^{{1\over2}\alpha' k_\perp^2}
\left[1-a(\eta M)_{\mu\nu}\tilde{\psi}_{-1/2}^\mu\psi_{-1/2}^\nu
+{\cal O}(a^2)\right]\left|\,0;-\sum k_\perp\right>.
\end{eqnarray}

Here $\sum k_\perp$ denotes the total transverse momentum of the other
vertex operators in the amplitude. The quantity $a=|z|$ now plays the
role of the hole radius in the bosonic discussion above. When
$a\rightarrow 0$, the boundary state moves to infinite distance, and
we recover a sphere amplitude with insertions of closed string
operators. Again there are poles in $k_\perp^2$ from the integration
around $a\approx0$. The first term is associated with the type II
tachyon and should never appear in appropriately GSO projected
amplitudes. The next term is the interesting one as it has a massless
pole. The corresponding closed string operator insertion is
\begin{equation}
(\eta M)_{\mu\nu}\tilde{\psi}_{-1/2}^\mu\psi_{-1/2}^\nu
\left|\,0;-\sum k_\perp\right>,
\end{equation}
which we recognize as part of the graviton vertex operator in the
ghost number $(-1,-1)$ picture. Of course, the insertion in (\ref{Asph})
was nothing but a part of the graviton vertex in the $(0,0)$ picture, so
we see again that there is a correspondence between the disk and the
sphere, exactly as in the bosonic case. We can further confirm that
the dependences on $\epsilon$ and $k_\perp^2$ are identical, and for
complete agreement we also verify that the numerical coefficients
are the same, since
\begin{equation}
{32\pi^3\over\alpha'\kappa^2}\cdot{1\over2\pi\alpha'}\cdot
{\alpha'\over2}\cdot\kappa^2\tau_3=
4\pi^2\tau_3\cdot{2\over\alpha'}.
\end{equation}
We have thus seen that the analysis for the superstring case goes
through exactly as the bosonic case, i.e., the full equation of motion
is not reproduced simply by demanding conformal invariance for
$\epsilon\rightarrow 0$.

\section{Conclusions and speculations}

\def\eps{\epsilon}
How should one interpret the prescription that we have presented above?
To take a
scale invariant prescription\cite{green},
and  introduce a scale dependence in order to cancel it
against another term, may appear to
be reckless, but has an illustrious precedent in string theory---the
Fradkin-Tseytlin\cite{ft,cfmp} term in the sigma model that is required for
obtaining consistent dilaton equations of motion.  It is
usual\cite{callan} to restate the sort of analysis we have gone
through in terms of a normalization scale dependence, phrasing the
argumentation in terms of anomalies in conformal invariance.
This is   {\it not}
possible  when one introduces a scale dependence in a {\it finite}
amplitude.  The `virtual anomaly' (a phrase due to J. Polchinski,
private communication) in scale invariance
introduced by the finite cutoff is required for spacetime diffeomorphism
invariance, much as finite counterterms in ordinary field theories are
required in many instances for preserving symmetries in the renormalized
theories.  The difference here is that we are concerned with
conformally invariant, hence finite, theories,
and the introduction of a scale in
such a theory requires that the scale have a physical interpretation.

It is also of interest  to recall that Banks and Martinec\cite{bm}
first suggested that
the Wilsonian renormalization group underlies string theory---of
course, the Wilsonian renormalization group automatically comes with
a finite cutoff.

If we do take the worldsheet cutoff seriously, we must
attempt to reconcile it with spacetime physics.  The following remarks
are speculative, but we will attempt to err on the conservative side in
our interpretation of the string phantasmagoria.

Clearly, $\eps$
must be very small in string perturbation theory for self-consistency,
so it is natural to conjecture an
inverse relation, $|\ln\eps|\sim g_{{\rm st}}^{{-\alpha}},$ with
$\alpha >0.$  This would imply $\eps \sim \exp\left(-C/g_{{\rm
st}}^{\alpha} \right),$ which would be consistent with perturbative string
theory.    With this kind of dependence
in the cutoff, we suggest the interpretation
that the vanishing of beta functions
is just the requirement  that the weak-coupling limit is
non-singular.
The worldsheet cutoff scale does not seem to
be directly related to Shenker's observations\cite{steve}.
Since strings are related to other $p$-branes in various dual formulations,
such cutoffs are presumably in the quantum field theories living on
the world-volumes of such objects as well.  For example, the
remarkable matrix discretization of membranes due to de Wit, Hoppe
and Nicolai\cite{dewit} may   be more directly relevant for M
theory than the continuum limit.   A finite worldsheet cutoff has a number of
physical consequences: (1) it implies a cutoff on the number of propagating
massive modes; (2) non-renormalization
theorems may receive corrections of order $\eps;$ and
(3)  the decoupling of the conformal mode needs to be reconsidered.

A rather more drastic cutoff, but still something difficult to see in
string perturbation theory, is a possible connection with the `stringy
exclusion principle' suggested by Maldacena and Strominger\cite{andy};
see also \cite{others}.  This principle suggests that the maximum
allowed occupation numbers of bosonic BPS particle modes grows in
inverse proportion to the coupling constant.  If it applies in
general\cite{andycomm}, though it is not entirely clear why it should, then
one
would expect a cutoff on the worldsheet that behaves as $\eps \propto
g_{{\rm st}}^{\alpha},$ which is considerably more drastic than the
cutoff suggested above.  Nevertheless, given the strong divergence of
string perturbation theory\cite{shenker},
it is not clear that such dependence of
the cutoff is immediately ruled out.  The spacetime length scale
corresponding to (\ref{length}),
$\ell \approx \sqrt{|\ln\eps|}$  still diverges,
but as $\sqrt{|\ln g_{{\rm }}|}.$  Further, even with this kind of
coupling constant dependence of the cutoff, the interpretation
of beta functions suggested above  still holds.

Gopakumar and Vafa\cite{gova}\ have recently suggested a novel
picture for the relation of the closed string sigma model description
to the hole description of D-branes.  They propose, in a topological
sigma model description, that a phase
separation mechanism might account for the appearance of holes, which
would be regions on the worldsheet where the fields would be frozen.
If this is indeed the case, then it might in fact be natural for the
size of the holes to be functions of the coupling constant, though
perhaps not in a topological string theory.

As we go to higher order in the string perturbation theory we would
expect similar effects to take place, with additional
complications due to regimes of colliding holes, etc.   We emphasize
that all the technical details of these calculations are similar to
those involved in the usual Fischler-Susskind mechanism.  While there
is no mathematical proof that the Fischler-Susskind mechanism is
consistent to all orders in perturbation theory, there has appeared no
evidence for any inconsistency in the physics literature.  Therefore,
while we make no claims to mathematical rigor, we are confident that
our scheme is on the same footing as the standard Fischler-Susskind
mechanism in this regard.

Starting at the annulus level, there are real divergences in
D0-brane amplitudes in the limit of shrinking handles. These
divergences are  related to the effects of the finite mass
D0-brane recoiling\cite{recoil}, and need to be taken into account
as well. It is interesting to note that through the current analysis
we have related the sphere and the disk amplitude, and thus the
formalism knows about the finite D0-brane mass. In particular, we can
use the full equations of motion, derived here from a world sheet
point of view, to calculate semiclassical scattering amplitudes
involving gravitons and D0-branes. The classical solutions for this
scattering process will necessarily involve a recoiling D0-brane,
recalling that the Einstein equation implies the geodesic equation in
general.

In summary, we have shown how the case of localized D-branes differs
from the open string case in that demanding conformal invariance on
the world sheet does not immediately imply the full spacetime
equations of motion. Instead the correct background metric and other
fields arise as a secondary effect from adding boundaries with
Dirichlet boundary conditions. We can still infer the full spacetime
equations of motion though, by introducing a cutoff by hand and
demanding that the leading cutoff dependence cancel between the
sphere and the disk amplitudes.  This prescription is consistent with T
duality.
It should be noted that a finite cutoff violates open--closed string
duality---this duality is, however, manifestly a perturbative
phenomenon.  There is {\it no} evidence that it continues
to have a physical significance at finite string coupling.
Indeed, at $g_{{{\rm st}}}>0,$ the  very validity of a string picture
is rather unclear, given the $2n!$ growth of string perturbation
theory\cite{shenker}.

\section*{Acknowledgments}

Conversations with T. Banks, E. D'Hoker,  D. Gross,
I. Klebanov, E. Martinec,
S. Shenker, A. Strominger, W. Taylor,   L. Thorlacius, and
especially C. Callan, W. Fischler and J. Polchinski,  are gratefully
acknowledged.
This work was supported in part by NSF grant PHY96-00258.  \O.T.
was supported in part by the Research Council of Norway.

\end{document}